\documentclass[prl,aps,twocolumn,psfig,showpacs]{revtex4}
\usepackage{epsfig}
\usepackage{graphicx}
\usepackage{amsmath}

\begin{document}

\title{Dynamic Structure in a Molten Binary Alloy by ab-initio 
Molecular Dynamics: Crossover from
Hydrodynamics to the Microscopic Regime.}

\author{D. J. Gonz\'alez, L. E. Gonz\'alez, J. M. L\'opez
and M. J. Stott$^{\S}$ }
\affiliation{Departamento de F\'\i sica Te\'orica, Universidad de Valladolid,
47011 Valladolid, SPAIN}
\affiliation{
$^{\S}$Department of Physics, Queen's University, Kingston, 
K7L 3N6 Ontario, CANADA}

\date{\today}

\begin{abstract}

The dynamic structure factor of the $^{7}$Li$_{\rm 0.61}$Na$_{\rm 0.39}$ liquid 
alloy at $T=590$ K has been calculated by {\it ab initio} molecular dynamics
simulations using 2000 particles. 
For small wavevectors, $0.15 \leq q/ $\AA$^{-1} \leq 1.6$, we find
clear side peaks in the partial dynamic structure factors. Whereas for
$q \leq 0.25$ \AA$^{-1}$ the peak frequencies correspond to the hydrodynamic
sound dispersion of the binary alloy, for greater $q$-values we obtain two
modes with phase velocities above and below the hydrodynamic sound. A smooth 
transition between hydrodynamic sound and the two collective modes is shown 
to take
place in the range  $0.25 \leq q/$ \AA$^{-1} \leq 0.35$.
The mass ratio in
this system, $m_{\rm Na}/m_{\rm Li} \approx 3$, is the smallest one so far 
for which the fast mode is observed. We also predict that inelastic X-ray 
scattering experiments would be able to detect the slow mode, and explain why
the inelastic neutron scattering measurements
 [P.R. Gartrell-Mills {\em et al}, 
Physica B {\bf 154}, 1 (1988)]
 do not show any of these modes.

\end{abstract}


\maketitle

Interest in the collective excitations in liquid binary mixtures was 
generated by the pioneering molecular dynamics (MD) study on the dynamics of 
the liquid Na-K alloy \cite{Jacucci&McDonald}. It was further stimulated by 
the MD results for liquid Li$_4$Pb, where a new, high-frequency mode,
supported by the light Li atoms only (the so-called ``fast sound") was 
identified \cite{Jacucci2}.
Kinetic theory calculations \cite{Bosse86} for two-component 
fluids with large atomic mass difference, confirmed the appearance of a 
collective mode at high frequencies, $\omega$, and wave numbers $q$ $\geq$  
$q_h$, where $q_h$ denotes the upper limit of the hydrodynamic regime. 
Moreover, this mode propagated with a phase velocity close to that of the 
pure light component, which was clearly greater than the hydrodynamic sound 
velocity, $c_{h}$, of the two-component system. Subsequently, calculations 
based on the revised Enskog theory for binary mixtures of hard spheres with
a large mass difference \cite{Campa888990} predicted two propagating 
collective modes in addition to hydrodynamic sound. One of these was 
identified with the fast sound while the other mode, which
had propagating phase velocities below $c_{h}$, was consequently 
termed ``slow sound".  
Theoretical  \cite{Campa888990,Westerhuijs,Bryk&Mryglod}, computer 
simulation \cite{Enciso95,Crevecoeur96,Perea98,Sampoli02} and experimental 
\cite{Crevecoeur96,Montfrooij89,Jong94,Alvarez98,Bafile01,Wegdam2} studies have 
investigated the existence and properties of these collective modes.
Experimental evidence for the fast sound has been obtained for systems such 
as H$_{2}$-Ar, H$_{2}$-Xe and  He-Xe  mixtures by light scattering 
\cite{Wegdam2}, and He-Ne, He-Ar and Li$_{4}$Pb by inelastic neutron
scattering (INS) \cite{Westerhuijs,Crevecoeur96,Perea98}. 
Note that all these systems except Li$_4$Pb are gas mixtures, whereas the 
latter is pseudoionic mixture, and that the mass ratio varies 
from around 33 for He-Xe to around 5 for He-Ne.

There has been discussion concerning the way in which the two collective
modes connect with the hydrodynamic one when $q_h$ is approached from above. 
Some theoretical and MD results for He-Ar and He-Ne mixtures have suggested 
that either the fast and slow sound merge into the hydrodynamic sound 
\cite{Enciso95,Crevecoeur96}, or that the fast sound disappears at $q_h$ 
and the slow sound merges into the hydrodynamic one, with a predicted value 
\cite{Westerhuijs,Enciso95,Crevecoeur96} for $q_h$ $\approx$ 0.07 \AA$^{-1}$.
A similar value for $q_h$ has recently been obtained for liquid Li$_4$Pb  
by both INS and by MD calculations \cite{Perea98,Alvarez98}. 
Recent MD calculations \cite{Sampoli02},
which followed the INS measurements of Bafile {\em et al}\  \cite{Bafile01},
for the dynamic structure of a He$_{\rm 0.77}$Ne$_{\rm 0.23}$
gas mixture at two densities, showed a clear crossover from hydrodynamics to
fast and slow modes at $0.2 \leq q/$ \AA$^{-1} \leq  0.5$ and gave $q_h \approx
0.2$, which is substantially greater than previous estimates.

The present communication addresses the previous questions while providing
further insight into the microscopic dynamics of the Li$_{\rm 0.61}$Na$_
{0.39}$ liquid alloy. This is a metallic system, with a rather small mass 
ratio $\approx $ 3, to which much theoretical and experimental attention has 
been devoted because of its strong phase separating tendencies. It has a 
consolute point at T $\approx $ 577 K and concentration $c_{\rm Li} 
\approx 0.64$, close to the ``zero alloy" composition ($c_{\rm Li} \approx 
0.61$) for which the total static structure factor, $S(q)$, reduces to the
concentration partial structure factor $S_{CC}(q)$, i.e. 
$S(q)=S_{CC}(q)/c_{\rm Li}c_{\rm Na}$, because of the negative scattering 
length of the isotope $^{7}$Li. The $S(q)$ has been measured at several 
temperatures and concentrations \cite{RupKnoll}, and INS  measurements 
\cite{McGreevy} have been performed for several temperatures at the zero 
alloy composition. The measured total dynamic structure factor, $S(q, 
\omega)$, in the explored $(q, \omega)$ region  decreased monotonically as 
a function of $\omega$, and side peaks, which would have been indicative of 
collective modes, were absent.

The present study has used the orbital-free {\it ab initio} molecular 
dynamics (OF-AIMD) method, where the forces acting on the nuclei are computed
from electronic structure calculations, based on density functional theory
(DFT), which are performed as the MD trajectory is generated. A simple liquid 
metallic alloy, A$_c$B$_{1-c}$, is treated as a disordered array of bare ions
interacting with the valence electrons through electron-ion potentials.
The total potential energy of the system can be written, within the 
Born-Oppenheimer approximation, as the sum of the direct ion-ion coulombic 
interaction energy and the ground state energy of the electronic system,  
$E_g[\rho_g(\vec{r}))]$. According to DFT, the ground state electronic 
density, $\rho_g(\vec{r})$ minimizes an energy functional which is given as 
the sum of the kinetic energy of independent electrons, $T_s[\rho]$, the 
classical Hartree electrostatic energy, $E_H[\rho]$, the exchange-correlation
energy, $E_{\rm xc}[\rho]$, for which we have adopted the local density 
approximation and finally, the electron-ion interaction energy, 
$E_{\rm ext}[\rho]$, for which we have used local ionic pseudopotentials 
constructed within DFT \cite{GGLS}. In the OF-AIMD approach 
\cite{HK,Perrot-MaddenLQRT} an explicit but approximate functional of the 
density is used for $T_s[\rho]$. Proposed functionals consist of  
the von Weizs\"acker term, $T_W[\rho(\vec{r})] = \frac18 \int d\vec{r} \, 
|\nabla \rho(\vec{r})|^2 /\rho(\vec{r})$, plus further terms chosen 
in order to reproduce correctly
some exactly known limits. Here, we have used a simplified 
version of the average density models, developed by Garc\'\i a-Gonz\'alez 
{\em et al}\ \cite{PGGs} in which $ T_s=T_W+T_{\beta}$, where

\begin{eqnarray}
T_{\beta} = \frac{3}{10} \int d\vec{r} \, \rho(\vec{r})^{5/3-2\beta}
\tilde{k}(\vec{r})^2 \\
\tilde{k}(\vec{r}) = (2k_F^0)^3 \int d\vec{s} \, k(\vec{s})
w_{\beta}(2k_F^0|\vec{r}-\vec{s}|)   \nonumber
\end{eqnarray}

\noindent
$k(\vec{r})=(3\pi^2)^{1/3} \;  \rho(\vec{r})^{\beta}$, $k_F^0$ is the Fermi 
wavevector for mean electron density $\rho_0$, and $w_{\beta}(x)$ is a 
weight function chosen so that both the linear response theory and the 
Thomas-Fermi limits are correctly recovered. Further details of this 
functional are given in reference [\onlinecite{GGLS}] and we merely note
that in the present simulations we have used $\beta=0.51$.

We have performed OF-AIMD simulations for the Li$_{\rm 0.61}$Na$_{\rm 0.39}$ 
liquid alloy at temperature $T=590$ K and number density $\rho$= 0.03218 
\AA$^{-3}$. The cubic simulation box contained 2000 particles. 
%
%
Given the ionic positions at time $t$, the electronic density is expanded in 
plane waves, and the energy functional is minimized with respect to the plane
wave coefficients yielding the ground state electronic density and energy.
The Hellmann-Feynman theorem is used to obtain the forces on the ions which, 
along with the Verlet leapfrog algorithm, are used to update the ionic 
positions and velocities. 
The timestep was $0.0025$ ps, the equilibration 
lasted for 10 ps and the calculation of properties was made averaging over 
another 60 ps. 
The cutoff energy for the plane wave expansion was $8.15$ Ryd,  
giving a basis of about 85000 planewaves. These choices allow a minimum 
$q$-value of $0.158$ \AA$^{-1}$ which will permit investigation of length 
and time scales covering a range which includes the hydrodynamic regime.

The total dynamic structure factor $S_T(q,\omega)$, which is directly related
to the intensity obtained in either an inelastic neutron (INS) or X-ray 
scattering (IXS) experiment, is a weighted average of the partials 
$S_{ij}(q, \omega)$, defined as

\begin{equation}
S_{ij}(q, \omega) = \dfrac{1}{2 \pi (N_{i}N_{j})^{1/2}}  
\int dt \;  e^{i \omega t} \langle \rho_i(\vec{q}, t) \cdot
\rho_j^*(\vec{q}, 0) \rangle
\end{equation}

\noindent where $N_i$ is the number of $i$-type particles,  the $\langle$....$
\rangle$ stands for the ensemble average, the asterisk denotes complex 
conjugation, $\rho_i(\vec{q}, t) = \sum_{l(i)=1}^{N_i} \; \exp [ i \vec{q} 
\cdot \vec{R}_{l(i)}(t) ] $ represents the  
density fluctuations of the $i$-th component
with wavevector $\vec q$\ , and $\vec{R}_{l(i)}(t)$ is the position of the
$i$-type particle $l$ at time $t$.
The $S_T(q,\omega)$ observed in an IXS experiment is given by

\begin{equation}
S_T^{\rm (IXS)}(q, \omega) = \sum_{i,j=1}^{2}  (c_i c_j)^{1/2}
\dfrac{f_i(q) f_j(q)}{ \langle f^2(q) \rangle}  S_{ij} (q, \omega)
\label{STixs}
\end{equation}

\noindent where $f_i(q)$ are the atomic scattering factors  and
$\langle f^2(q)\rangle$ =$ \sum_{i=1}^{2} c_i$ $f_i^{2}(q) $.
Correspondingly, in an INS experiment

\begin{eqnarray}
\langle b^2 \rangle S_T^{\rm (INS)}(q, \omega) & =
 & \sum_{i=1}^{2} (\langle b_i^2 \rangle - \langle b_i \rangle^2) c_i S_i^s(q, \omega) \nonumber  \\
&
 + & \sum_{i,j=1}^{2}  (c_i c_j)^{1/2} \langle b_i \rangle \langle b_j \rangle S_{ij}(q, \omega)  
\label{STins}
\end{eqnarray}

\noindent where the $S_i^s(q, \omega)$ are the self-dynamic structure 
factors, $\langle b_i \rangle$ is the coherent scattering length, 
$4\pi\langle b_i^2 \rangle$ is the total scattering cross section and 
$\langle b^2 \rangle$ = $ \sum_{i=1}^{2}$ $c_i \langle b_i^2 \rangle$ is the 
average cross section per atom.

\begin{figure}
\mbox{\epsfig{file=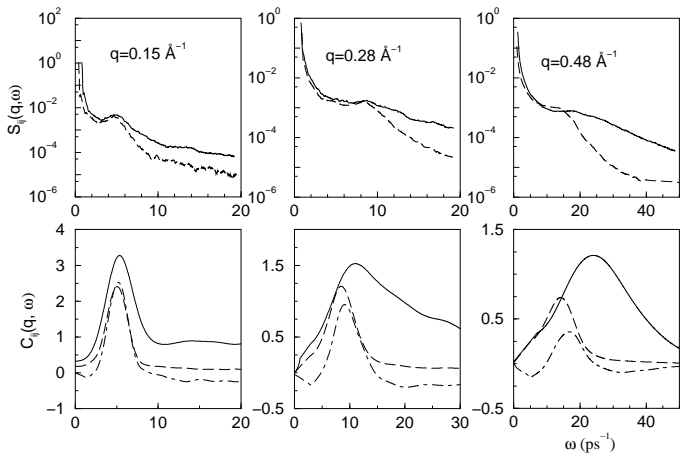, angle=-90 ,width=90mm } }
\caption{OF-AIMD results, at three $q$-values, for the liquid Li$_{\rm 0.61}$Na$_{\rm 0.39}$ at $T$=590 K.
Upper frame:  Partial dynamic structure factors (logarithmic scale). $S_{\rm LiLi}(q, \omega)$ (full lines) and
 $S_{\rm NaNa}(q, \omega)$ (dashed lines).
Lower frames: Partial longitudinal currents: $C_{\rm LiLi}(q, \omega)$ (full lines),
 $C_{\rm NaNa}(q, \omega)$ (dashed lines)  and  $C_{\rm LiNa}(q, \omega)$ (dot-dashed lines). }
\label{sckwij}
\end{figure}

The results for $S_{\rm LiLi}(q, \omega)$ and $S_{\rm NaNa}(q, \omega)$ are 
shown in fig.\  \ref{sckwij} for three representative $q$-values. At the 
smallest $q$-value allowed by the simulations, $q$= 0.158 \AA$^{-1}$, both 
$S_{\rm LiLi}(q, \omega)$ and $S_{\rm NaNa}(q, \omega)$ show clear Brillouin 
peaks at a common frequency, $\omega_B$, which is the typical behaviour in the 
hydrodynamic region and indicates a propagating sound mode with a hydrodynamic
velocity $c_h$ = $\omega_B$/$q$ $\approx$ 3000 m/s. $S_{\rm LiLi}(q, 
\omega)$ exhibits Brillouin peaks for the other two $q$-values of  fig.\  
\ref{sckwij} and continues to show side peaks or shoulders up to $q$ 
$\approx$ 1.5 \AA$^{-1}$, whereas $S_{\rm NaNa}(q, \omega)$ has only a 
shoulder for $q=0.48$ \AA$^{-1}$ and this feature disappears at $q$ $\approx$ 
0.9 \AA$^{-1}$. In contrast, the partial longitudinal 
currents, $C_{ij}(q, \omega)$= $\omega^{2}$ $S_{ij}(q, \omega)$, show 
clear peaks at any $q$-value. At the 
lowest $q$-value, the frequencies of the maxima in the $C_{ij}(q, \omega)$ 
coincide with those of the Brillouin peaks, but at larger $q$ they are 
somewhat higher because of the contribution from the damping terms 
\cite{Enciso95}. Moreover, as $q$ increases $C_{12}(q, \omega)$ diminishes
pointing to a progressive decoupling of the motions of two species.

\begin{figure}
\mbox{\epsfig{file=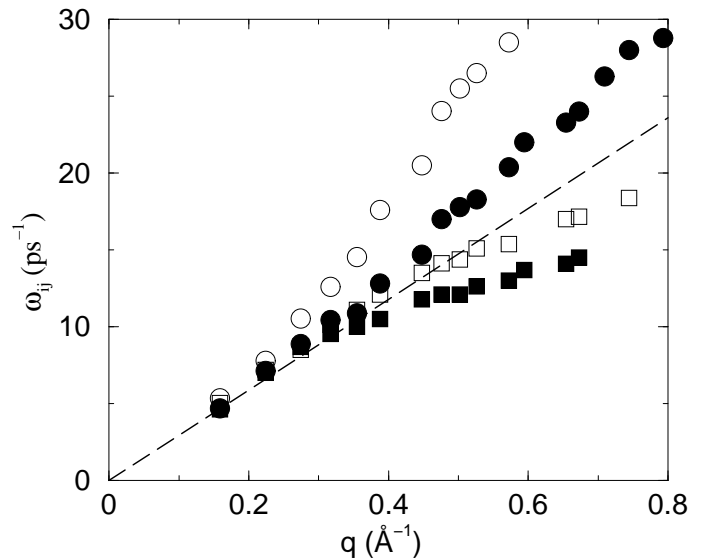, angle=-90 ,width=90mm } }
\caption{OF-AIMD results for the dispersion curves of the collective modes in
$S_{\rm LiLi}(q, \omega)$ and
$S_{\rm NaNa}(q, \omega)$ (full circles and squares respectively). The positions of the maxima of
$C_{\rm LiLi}(q, \omega)$ and $C_{\rm NaNa}(q, \omega)$ are also shown
(open circles and squares respectively).}
\label{dispersion}
\end{figure}

The dispersion curves $\omega_{\rm LiLi}(q)$  and $\omega_{\rm NaNa}(q)$ of 
the inelastic peaks in  $S_{\rm LiLi}(q, \omega)$ and $S_{\rm NaNa}(q, 
\omega)$  respectively, are shown in fig.\  \ref{dispersion}. Up to $q$ 
$\approx$ 0.3 \AA$^{-1}$, $\omega_{\rm LiLi}(q)$ and $\omega_{\rm NaNa}(q)$ 
show the same linear behaviour of hydrodynamic collective excitations 
propagating with speed $c_h$ $\approx$ 3000 m/s.  Above this $q$-value, the 
dispersion curve splits into two branches signalling the onset of dynamic 
decoupling of the species, and giving rise to fast and slow sound modes in 
the alloy. The fast mode 
involves the Li particles only, with a 
phase velocity  $c_{\rm fast}$ =$\omega_{\rm LiLi}(q)$ /$q $ $\approx$ 3800 
m/s, whereas the slow mode phase velocity is $c_{\rm slow}$ = $\omega_{\rm 
NaNa}(q)$ /$q $ $\approx$ 2100 m/s, at $q$=0.65 \AA$^{-1}$.  We have also 
performed OF-AIMD simulations for pure Li at the same temperature and total 
number density as the alloy, and the calculated sound velocity, 3900 m/s, is 
very close to the velocity of the fast mode. Following usual practice 
\cite{Enciso95}, we have also obtained dispersion curves derived from the 
maxima in $C_{ij}(q, \omega)$, and these are also shown 
in fig.\  \ref{dispersion}.  The frequencies for both Li and Na
are larger than $\omega_{\rm LiLi}(q)$ and  $\omega_{\rm NaNa}(q)$, and the
earlier onset of the branching suggests that the departure from hydrodynamics
begins already at $q$ $\approx$ 0.15 \AA$^{-1}$.
Moreover, the higher slope obtained for the Li leads to a larger estimation 
for the phase velocity of the fast mode ($c_{\rm fast}$ $\approx$ 4600 m/s).

\begin{figure}
\mbox{\epsfig{file=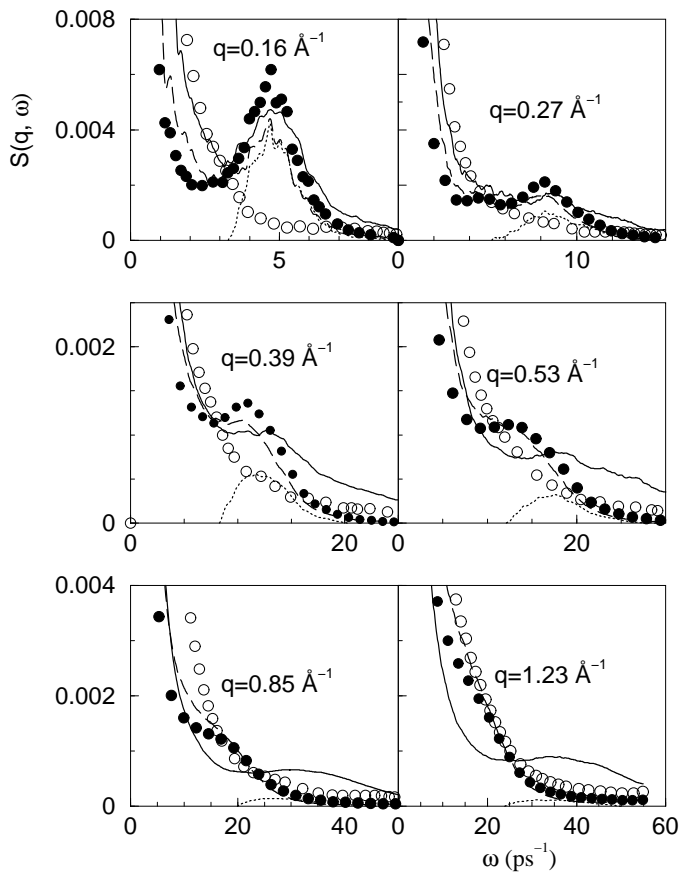, angle=-90 ,width=90mm } }
\caption{OF-AIMD results for the total dynamic structure factors in
liquid Li$_{\rm 0.61}$Na$_{\rm 0.39}$ at $T$=590 K.  Open circles: $S_T^{\rm (INS)}(q, \omega)$.
Full circles: $S_T^{\rm (IXS)}(q, \omega)$. For comparison, the  partials $S_{\rm LiLi}(q, \omega)$ (full line),
$S_{\rm NaNa}(q, \omega)$ (dashed line)  and $S_{\rm LiNa}(q, \omega)$ (dotted line)
are also included. }
\label{Stnx}
\end{figure}

Now we consider the possible experimental observation of these collective modes.
As already mentioned, the INS  measurements \cite{McGreevy} gave a
total structure factor with no side peaks as a function of $\omega$.
We have calculated  $S_T^{\rm (INS)}(q, \omega)$ using eqn.\  (\ref{STins}) 
and data from our simulations. The results illustrated in fig.\  \ref{Stnx} 
do not show side peaks in agreement with the experiment, and despite the 
appearance of collective modes in the simulation. For an understanding 
of this anomaly we have also plotted in fig.\  \ref{Stnx} the partials 
$S_{ij}(q, \omega)$. We see that $S_{\rm LiNa}(q, \omega)$ also exhibits a 
peak at a frequency between that of $S_{\rm LiLi}(q, \omega)$  and $S_{\rm 
NaNa}(q, \omega)$. In the weighted average of eqn.\  (\ref{STins}), the 
negative scattering length of the $^{7}$Li isotope ($\langle$$b_{\rm 
Li}$$\rangle$=--0.22 $\times$10$^{-12}$ cm) gives rise to a negative 
contribution from the partial  $S_{\rm LiNa}(q, \omega)$ which balances 
the contributions from the peaks in $S_{\rm LiLi}(q, \omega)$  and $S_{\rm 
NaNa}(q, \omega)$. In contrast, we predict that side peaks would be observed 
in an IXS measurement because the atomic form factors involved in the 
average in eqn.\  (\ref{STixs}) always take positive values. Moreover, 
the peaks would be those due to the slow sound because the prefactor of 
$S_{\rm NaNa}(q, \omega)$ is about twenty times larger than that of $S_{\rm 
LiLi}(q, \omega)$ and twice that of $S_{\rm LiNa}(q, \omega)$, with the 
overall result that the the total $S^{\rm (IXS)}_T(q, \omega)$ is mainly 
controlled by $S_{\rm NaNa}(q, \omega)$.  

In conclusion, our simulations of the 
Li$_{\rm 0.61}$Na$_{\rm 0.39}$ liquid alloy 
show well defined collective modes analogous to those already found in other
binary fluids. However, this system has the smallest mass ratio 
considered so far and is the 
first metallic system for which ``fast sound" appears.
The results for the INS total dynamic structure factor explain the
reason for the failure to detect collective excitations in the INS experiment
by Gartrell-Mills {\it et al} \cite{McGreevy}. However, we predict that these 
excitations, particularly the slow sound mode, would be observed in an IXS 
experiment. 

In order to specify where the transition from hydrodynamics to kinetic regime 
takes place, we have found that it is important to specify 
which magnitude is used. In 
particular, partial currents depart from hydrodynamic behavior 
for smaller $q$-values 
than the corresponding partial structure factors, and this must be taken 
into account in discussions
of the onset of ``fast sound".

Finally, we have shown that is now possible to 
perform {\it ab-initio} molecular 
dynamics simulations to study the dynamic properties of fluids. The OF-AIMD 
method, which employs the electron density as the basic variable, gives an 
approximate treatment of the electron kinetic energy, but within this limitation
allows the simulation of large samples for long runs. 
These are basic requirements 
for performing reliably the ensemble averages that 
define most dynamic properties 
in liquids, as well as for studying the ordering 
tendencies in multicomponent systems.

This work has been supported by the Junta de Castilla y Le\'on (VA 073/02) and
CICYT (PB98-0368-C02).  DJG acknowledges the
University of Valladolid for financial support.
MJS acknowledges the support of the NSERC of Canada.

\end{document}